\pdfoutput=1
\documentclass[pdflatex,sn-mathphys-num]{sn-jnl}   

\usepackage{graphicx}%
\usepackage{multirow}
\usepackage{amsmath,amssymb,amsfonts}%
\usepackage{amsthm}%
\usepackage{mathrsfs}
\usepackage[title]{appendix}%
\usepackage{xcolor}%
\usepackage{textcomp}%
\usepackage{manyfoot}%
\usepackage{booktabs}%
\usepackage{algorithm}%
\usepackage{algorithmicx}%
\usepackage{algpseudocode}%
\usepackage{listings}%
\usepackage{pifont}
\newcommand{\cmark}{\ding{51}}
\newcommand{\xmark}{\ding{55}}
\usepackage{diagbox}
\usepackage{url}

\theoremstyle{thmstyleone}%

\theoremstyle{thmstyletwo}%
\theoremstyle{thmstylethree}%

\begin{document}

\title[Article Title]{Rongzai agent: A Large Language Model-Based Autonomous Assistant for Rietveld Refinement of Neutron Diffraction Data}

\author[1]{\fnm{Qingmeng} \sur{Li}}\email{liqm@ihep.ac.com}
\equalcont{These authors contributed equally to this work.}

\author[1,2]{\fnm{Hao} \sur{Wang}}\email{wanghao96@ihep.ac.com}
\equalcont{These authors contributed equally to this work.}

\author[1]{\fnm{Dongbo} \sur{Xiong}}\email{xingdb@ihep.ac.com}
\equalcont{These authors contributed equally to this work.}

\author[1,2]{\fnm{Jiajun} \sur{Zhong}}\email{zhongjiajun@ihep.ac.cn}

\author[1,2]{\fnm{Wenhai} \sur{Ji}}\email{jiwh@ihep.ac.cn}

\author[1,3]{\fnm{Hao} \sur{Hu}}\email{huhao@ihep.ac.cn}
\author[1]{\fnm{Yiyu} \sur{Zhang}}\email{zhangyiyu@ihep.ac.cn}
\author[1]{\fnm{Bolun} \sur{Zhang}}\email{zhangbolun@ihep.ac.cn}
\author[1,3]{\fnm{Hong} \sur{Wang}}\email{wanghong@ihep.ac.cn}
\author[1]{\fnm{Yongfeng} \sur{Zhu}}\email{zhuyf@ihep.ac.cn}

\author*[1,2]{\fnm{Rong} \sur{Du}}\email{durong@ihep.ac.cn}

\author*[1,3,4]{\fnm{Zhengde} \sur{Zhang}}\email{zdzhang@ihep.ac.cn}

\author*[1,3,4]{\fnm{Fazhi} \sur{Qi}}\email{qfz@ihep.ac.cn}

\author*[1,2,3]{\fnm{Junrong} \sur{Zhang}}\email{jrzhang@ihep.ac.cn}

\affil[1]{\orgdiv{Institute of High Energy Physics}, \orgname{Chinese Academy of Sciences}, \orgaddress{\street{Yuquan Road}, \city{Shijingshan District}, \postcode{100049}, \state{Beijing}, \country{People’s Republic of China}}}

\affil[2]{\orgname{Spallation Neutron Source Science Center}, \orgaddress{\street{Zhongziyuang Road}, \city{Dongguan}, \postcode{523803}, \state{Guangdong}, \country{People’s Republic of China}}}

\affil[3]{\orgname{National High Energy Physics Science Data Center}, \orgaddress{\street{Yuquan Road}, \city{Shijingshan District}, \postcode{100049}, \state{Beijing}, \country{People’s Republic of China}}}

\affil[4]{\orgname{University of Chinese Academy of
Sciences}, \orgaddress{\street{Yuquan Road}, \city{Shijingshan District}, \postcode{100049}, \state{Beijing}, \country{People’s Republic of China}}}


\abstract{Neutron diffraction (ND) is an indispensable technique for determining atomic positions (especially light elements) and thus serves as a critical probe for revealing microscopic structures in materials science. However, traditional Rietveld refinement of ND data relies heavily on manual operation of specialized software, which is time‑consuming, labor‑intensive, and highly dependent on user expertise, severely hindering automated analysis. The automation of Rietveld refinement has long been a long‑standing and challenging problem in crystallography. To address this challenge, this paper presents the Dr.Sai-Rongzai agent, an autonomous refinement assistant based on a large language model (LLM), a specialist knowledge base, and the GSAS‑II refinement engine, achieving for the first time an intelligent refinement that integrates knowledge‑driven decision‑making. The agent accomplishes a fully automated workflow from natural language task parsing to autonomous decision‑making, execution of refinement strategies, and report generation. Evaluation on five representative samples shows that the Rongzai agent achieves lower $R_{\text{wp}}$ values than human specialists on three samples (2.88\% vs. 4.42\%, 5.06\% vs. 5.40\%, 7.60\% vs. 9.00\%), while on the other two samples its results are very close to those of the specialists. The agent is currently deployed at the China Spallation Neutron Source (CSNS) and is open for external user registration, providing an intelligent and user friendly analytical tool for materials research. This work fully leverages the cutting‑edge advantages of LLM, offers a new path to solve the long‑standing problem of automated refinement, takes a key step toward intelligent and fully automated crystallographic analysis, and holds great potential to accelerate AI for Science discoveries in neutron‑based materials characterization.}

\keywords{Neutron diffraction, Rietveld refinement, Rongzai agent, large language model, AI for Science}

\maketitle

\section{Introduction}\label{sec1}

Neutron powder diffraction (NPD) is a central probe of crystal and magnetic structure, yet converting diffraction patterns into reliable structural models remains a major practical bottleneck in materials research \cite{haberl_advancing_2023,pomjakushin_topological_2023,vershinina_comparative_2023}. The difficulty is intrinsic: powder diffraction compresses three-dimensional structural information into a one-dimensional profile \cite{de_wolff_determination_1957,mighell_geometrical_1975}. In principle, this inverse problem can be solved by Rietveld refinement, which fits a physically parameterized model to the full diffraction pattern \cite{rodriguez-carvajal_recent_2001,toby_gsas-ii_2013}. In practice, however, success depends not only on optimization, but on choosing an appropriate refinement path. Because structural, instrumental and microstructural parameters are often strongly coupled, they must be released in a physically sensible order, with repeated adjustments based on intermediate outcomes \cite{toby_simple_2024,young_rietveld_1993}. Rietveld refinement is therefore not merely a nonlinear least-squares problem, but a scientific decision process carried out under physical constraints and uncertainty. This dependence on expert judgment has become a critical bottleneck for high-throughput and parametric diffraction studies, where large numbers of datasets must be analysed rapidly and consistently \cite{biwer_spotlight_2025}.

A wide range of automated refinement methods has been developed to address this challenge, including fixed workflows, rule-based expert systems, reinforcement learning, and Bayesian or global optimization strategies \cite{tian_srrietveld_2013,cui_autofp_2015,feng_method_2019,aimi_development_2019,ozaki_automated_2020,biwer_spotlight_2025}. Although these approaches have advanced the field, they have not resolved its core limitation: robust refinement requires context-aware, sequential, and corrective decision-making. Predefined workflows are brittle when confronted with unexpected refinement behaviour; rule-based systems are constrained by manually encoded knowledge; and search-based methods can be computationally intensive while still lacking the capacity to interpret intermediate failures and revise strategy accordingly. As a result, general-purpose autonomous refinement that is both reliable and practical on real experimental data remains out of reach.

Recent progress in large language model (LLM)-based agents suggests a different route. By integrating reasoning, planning and tool use in a closed feedback loop, agents can perform tasks that require iterative judgment rather than one-shot prediction \cite{ai_scivisagentbench_2026,wu_omniflow_2026}. This is precisely the structure of Rietveld refinement: a successful analysis requires evaluating partial fits, deciding what to refine next, identifying when a strategy has become unstable or unphysical, and recovering accordingly. The challenge, therefore, is not simply to automate software execution, but to operationalize expert refinement logic in an autonomous, tool-grounded system.

\begin{figure}[htbp]
    \centering
    \includegraphics[width=0.8\textwidth]{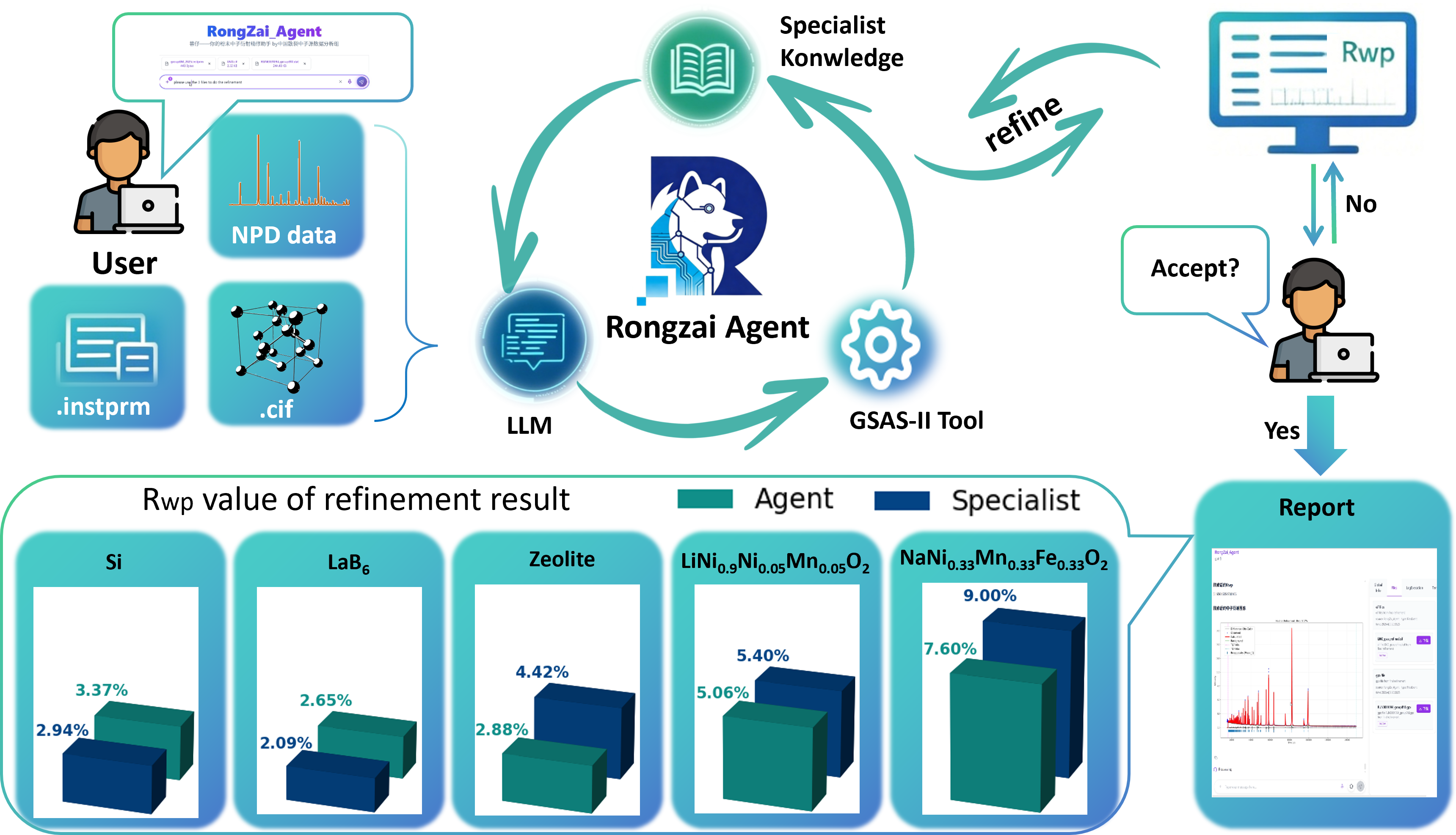}
    \caption{Schematic of the Rongzai agent workflow. The agent integrates a LLM, specialist knowledge, and a GSAS-II tool to perform automated, iterative NPD refinement, achieving $R_{\text{wp}}$ results competitive with or better than human experts.}
    \label{fig:agentWorkflow}
\end{figure}

Here we present Dr.Sai-RongZai, the first intelligent agent for autonomous NPD Rietveld refinement that achieves knowledge-driven decision-making by integrating a LLM with specialist refinement knowledge. The work builds on Open Dr.Sai (https://github.com/hepai-lab/drsai), a scientific AI agent framework developed by our team. The agent combines an LLM, specialist refinement knowledge, and a custom GSAS-II execution interface within an iterative decision–execution–evaluation loop (Fig.~\ref{fig:agentWorkflow}). Given diffraction data and structural inputs, it plans refinement actions, executes them, interprets intermediate outcomes, and updates its strategy until convergence. Across multiple representative samples, RongZai agent achieves refinement quality comparable to, and in several cases better than, that of human specialists, as measured by $R_{\text{wp}}$.

These results establish an agentic paradigm for diffraction refinement. Rather than treating automation as fixed workflow execution or expensive parameter search, our approach frames refinement as adaptive scientific reasoning grounded in professional software and domain knowledge. More broadly, this work shows that LLM agents can serve not only as interfaces to scientific tools, but as autonomous operators of expert analytical workflows, opening a practical path toward scalable data analysis in neutron scattering and related fields.

\section{Development method of Rongzai Agent}
\subsection{The architecture of Rongzai Agent}

Rongzai Agent is developed based on the Open Dr.Sai framework. The Open Dr.Sai framework introduces six specialized modular components addressing critical scientific computing requirements: perception modules for scientific data ingestion, long-term memory management with RAGFlow integration, scientific tool execution supporting MCP/OpenAPI/HepAI Worker protocols, sophisticated state management with human-in-the-loop capabilities, long-running task monitoring with multi-process architecture, and a dynamic agent skill system using progressive disclosure design, offering comprehensive support for agent development.

\begin{algorithm}[htbp]
\caption{Feedback-Driven Self-Loop Architecture}\label{alg:feedback-loop}
\begin{algorithmic}[1]
\State context.add(userPrompt)
\While{True}
    \State $output \gets \text{LLM}(systemPrompt, context)$
    \State context.add($output$)
    \If{$output[\text{``require''}] == \text{``refine init''}$}
        \State $initializationResult \gets \text{refineInitTool}(output[\text{``raw files''}])$
        \State context.add($initializationResult$)
    \ElsIf{$output[\text{``require''}] == \text{``refine''}$}
        \State $refineResult \gets \text{refineTool}(output[\text{``gpx file''}], output[\text{``refine parameters''}])$
        \State context.add($refineResult$)
    \ElsIf{$output[\text{``require''}] == \text{``refine pause''}$}
        \State $refineReportTemp \gets \text{refinePauseTool}(output)$
        \State \Return
    \ElsIf{$output[\text{``require''}] == \text{``refine finish''}$}
        \State $refineReportFinal \gets \text{refineFinishTool}(output)$
        \State \Return
    \EndIf
\EndWhile
\end{algorithmic}
\end{algorithm}

The core of Rongzai Agent consists of three components: a LLM, the GSAS-II refinement engine, and an expert knowledge base. The LLM is responsible for task decomposition and refinement decision-making. GSAS-II, after secondary development, serves as an automated refinement tool that executes specific tasks. The expert knowledge base encodes the refinement experience accumulated by domain experts into system prompts, guiding the LLM to make professional decisions. The agent employs a "Feedback-Driven Self-Loop" architecture, as shown in Alg.~\ref{alg:feedback-loop}, achieving dynamic iterative optimization of refinement parameters through continuous feedback of multimodal information, including refinement indices and fitting plots. The LLM interacts with the GSAS-II tool through four dedicated code blocks: refine init (initializing the refinement project), refine (executing refinement steps), refine pause (pausing and waiting for user instructions), and refine finish (completing the task and outputting the report), thereby completing the entire workflow from initialization to final report generation. After the user provides instructions and input files, the system automatically performs iterative refinement until the results are acceptable. Throughout the entire task, contextual information is fully preserved, ensuring that the LLM can follow long-term strategies.

\subsection{The prompts design for LLM}
\begin{figure}[htbp]
    \centering
    \includegraphics[width=0.9\textwidth]{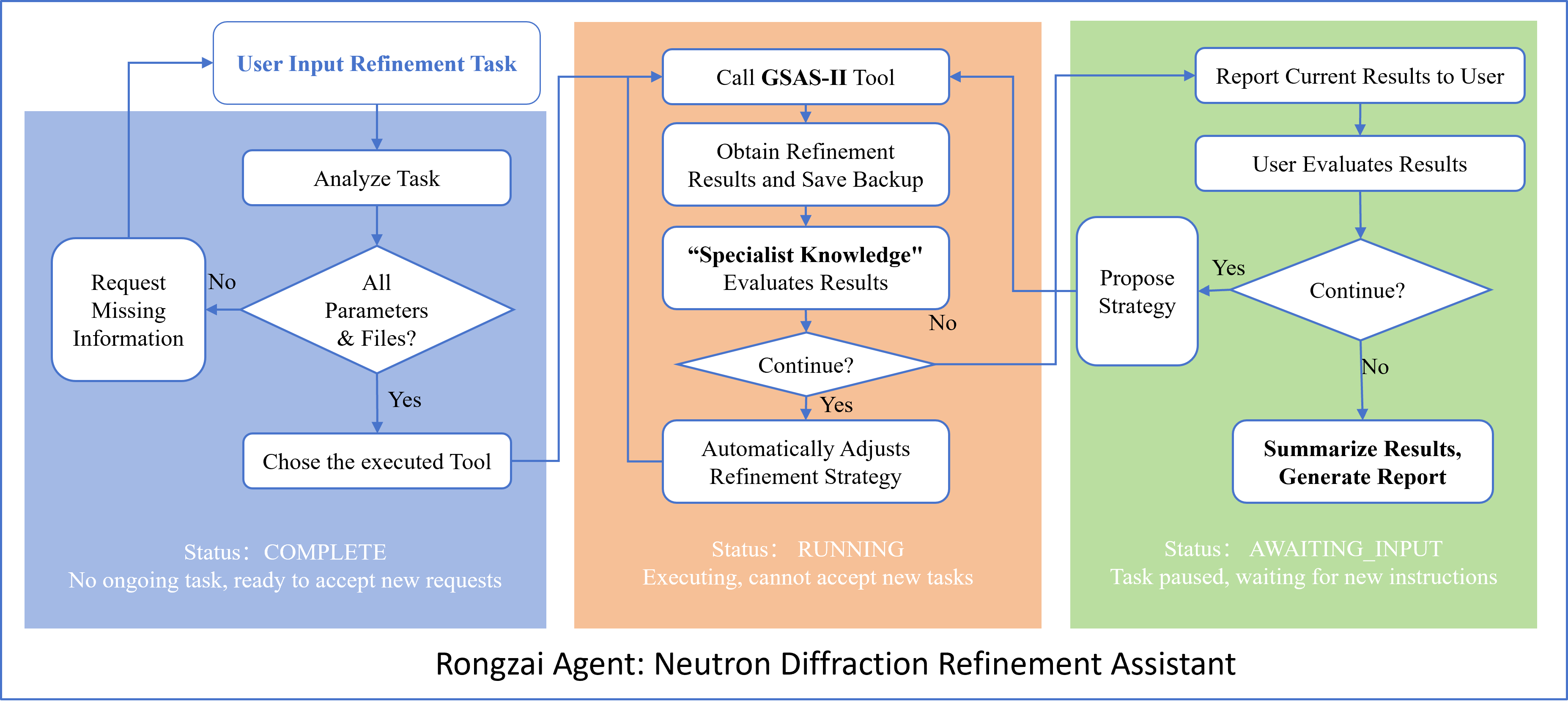}
    \caption{Flowchart of the prompt design and refinement process for Rongzai agent. The diagram illustrates the linear iterative refinement workflow driven by a three-state machine, incorporating expert knowledge encoding, parameter adjustment, backup/rollback mechanisms, and standardized report generation.}
    \label{fig:llm-prompt}
\end{figure}

In LLM application development, prompt design is essential for building effective agents. This section illustrates the principles and best practices of prompt design in the Rongzai agent. First, the LLM is clearly defined as a "neutron diffraction refinement assistant," establishing its professional boundaries. A three-state machine is then constructed to manage idle, executing, and waiting modes, ensuring appropriate responses based on the current state. The refinement process follows a linear iterative model. The user provides one instruction and three files: a diffraction pattern (.gsa), instrument parameters (.instprm), and crystal information (.cif). The LLM generates a GSAS-II project file (.gpx) to start the task. As shown in Fig.~\ref{fig:llm-prompt}, the process enters the COMPLETE state upon the first request, checks data completeness, then switches to RUNNING to execute refinement. After each round, results are evaluated. If improvement is possible, the system saves a backup and adjusts parameters using embedded expert knowledge. When criteria are met or a local optimum is reached, it switches to AWAITING\_INPUT, presenting results and suggestions to the user. The user may roll back to a previous backup or adjust the strategy. Once accepted, the agent outputs a standardized report with clear markers for downstream parsing. This state-machine and backup mechanism ensures flexibility and traceability. This multi-layered approach, which includes role definition, state management, specialist knowledge, and specific tool handling, forms a reusable framework for prompt design, offering systematic guidance for building domain-specific LLM agents.

\subsection{The GSAS-II refinement tool}
GSAS-II is a comprehensive open-source crystallographic analysis software package capable of handling the complete Rietveld refinement workflow. Based on GSAS-II, we have developed an automated refinement tool specifically designed for NPD, achieving full-process automation from project creation, parameter configuration, and refinement execution to result exportation via remote API calls. The tool adopts a FastMCP server architecture, supporting remote calls through the SSE protocol and asynchronous concurrent processing of multiple tasks. Core functionalities include automatic project creation and data import, intelligent cleaning of anomalous values, background parameter optimization, and flexible configuration of various refinement parameters. The accessible refinement parameters are listed in Table~\ref{tab1}, covering background, unit cell, scale factor, instrument parameters, atomic positions, sample displacement, crystallite size, microstrain, phase fraction, and other categories. Among these, some parameters can be either refined or manually set, providing flexible adjustment space for the agent. Upon completion of refinement, the system automatically performs data consistency verification, generates high-resolution plots (including observed values, calculated values, background, and Bragg peak positions), and outputs CSV data, structural parameter text files, and standard CIF format files. This tool significantly improves the efficiency and reproducibility of neutron powder diffraction data refinement.

\begin{table}[htbp]
\caption{Refinement parameters configurable by the Rongzai agent.}\label{tab1}
\centering
\begin{tabular}{@{}llcc@{}}
\toprule
Parameter & Description & Refinable & Settable \\
\midrule
cycle & Number of refinement cycles & \xmark & \cmark \\
background & Background parameters & \cmark & \xmark \\
scale & Scale factor & \cmark & \xmark \\
cell & Unit cell parameters & \cmark & \cmark \\
absorption & Absorption correction & \cmark & \xmark \\
alpha & Profile parameter alpha & \cmark & \xmark \\
beta & Profile parameters beta-0, -1, -q & \cmark & \xmark \\
sig & Profile parameters sig-0, -1, -2, -q & \cmark & \xmark \\
X & Profile parameter X & \cmark & \xmark \\
Y & Profile parameter Y & \cmark & \xmark \\
Z & Profile parameter Z & \cmark & \xmark \\
atom positions & Atomic coordinates & \cmark & \xmark \\
atom occupation & Site occupancy & \cmark & \xmark \\
atom uiso & Isotropic displacement parameter & \cmark & \cmark \\
size & Crystallite size & \cmark & \xmark \\
mustrain & Microstrain & \cmark & \cmark \\
phase fraction & Phase fraction & \cmark & \xmark \\
\botrule
\end{tabular}
\end{table}

\section{Results and discussion}\label{sec2}
\subsection{Experimental setup}
To objectively evaluate the refinement capability and efficiency of the Rongzai agent in practical applications, a systematic comparative experiment is designed. The experiment tests the refinement performance of the Rongzai agent on five materials using four different base models. The results are then compared with those of two human specialists in terms of refinement cycles, refinement quality, and time consumption. The four base models employed for Rongzai agent are GPT-4.1 and GPT-5 from OpenAI (https://openai.com), MiniMax-m2.7 from MiniMax (https://www.minimaxi.com), and Scienceone-s1-base-ultra from the ScienceOne team (https://www.scienceone.cn). Among the two human specialists, Specialist~I is a member of the CSNS data analysis group, specializing in structure determination and refinement using powder diffraction patterns. Specialist~II is an instrument scientist at the high-resolution diffraction instrument (TREND) beamline, specializing in neutron diffraction techniques with extensive experience in Rietveld refinement. It should be noted that although the Rongzai agent allows users to provide refinement strategy suggestions during the refinement process, all refinement results reported in this paper are obtained without any user intervention to ensure a fair comparison.

\subsubsection{Materials and data collection}
Five experimental patterns are used to evaluate the performance of the Rongzai agent: Si, LaB\(_6\), zeolite, Lithium-ion battery cathode (LiNi\textsubscript{0.9}Ni\textsubscript{0.05}Mn\textsubscript{0.05}O\textsubscript{2}) \cite{li_constructing_2026} and Sodium-ion battery cathode (NaNi\textsubscript{0.33}Mn\textsubscript{0.33}Fe\textsubscript{0.33}O\textsubscript{2}) \cite{xie_probing_2018} . These materials cover a diverse range of structural complexities, from simple crystalline standards to practical energy materials, providing a comprehensive testbed for the agent's capability. The initial crystal structures of these materials have been previously determined. The Si pattern is acquired using the Multi-Physics Instrument (MPI) of CSNS, while the remaining four patterns are acquired using the TREND of CSNS.

\subsubsection{Evaluation metrics}

The weighted profile R-factor ($R_{\text{wp}}$) is used to characterize the relative goodness of fit between the experimental data and the refined pattern. A lower $R_{\text{wp}}$ value indicates a better refinement result for the same data. In GSAS-II, the $R_{\text{wp}}$ is reported as "\text{wR}" in the output and follows the standard definition shown in Eq. ~\ref{eq1}.
\begin{equation}
R_{\text{wp}} = \sqrt{ \frac{ \sum_{i} w_i (y_i^{obs} - y_i^{calc})^2 }{ \sum_{i} w_i (y_i^{obs})^2 } }\label{eq1}
\end{equation}
The $y_i^{obs}$ and the $y_i^{calc}$ are the experimental and calculated pattern intensities, respectively. The corresponding weight $w_i$ is assigned to each data point, typically derived from counting statistics based on Poisson distribution, where $w_i \approx 1/y_i^{obs}$.   

\subsection{Iterative refinement process comparison}
\begin{figure}[H]
    \centering
    \includegraphics[width=1.0\textwidth]{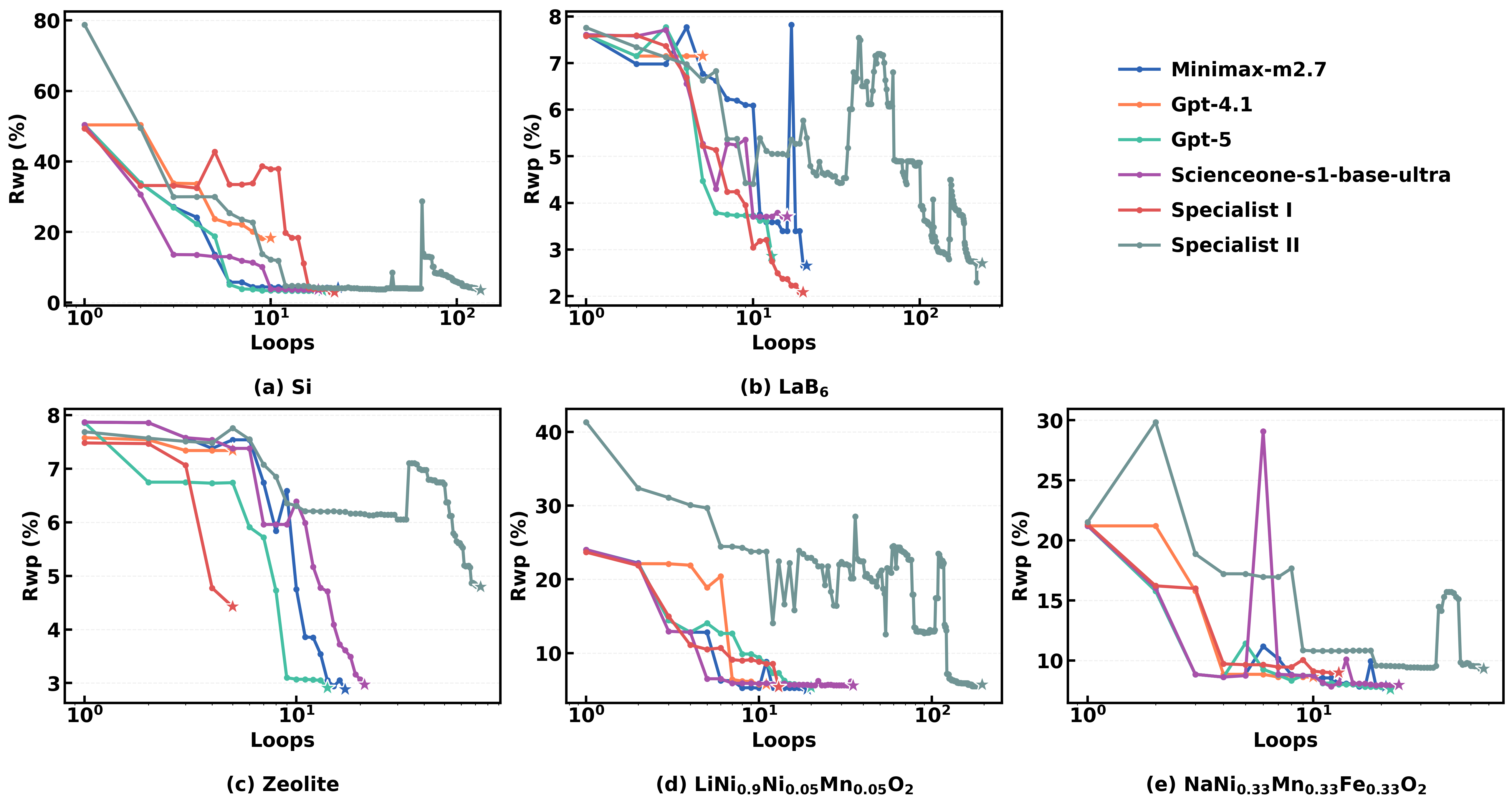}
    \caption{Evolution of $R_{\text{wp}}$ values with iteration number during the refinement of five samples. The four base models of Rongzai agent (MiniMax-m2.7, GPT-4.1, GPT-5, Scienceone-s1-base-ultra) and two specialists (Specialist I, II) are shown in blue, orange, green, purple, red, and gray, respectively, with final results marked by same-colored stars.}
    \label{fig:rwp change}
\end{figure}

To compare the convergence behavior of the Rongzai agent and human specialists during the refinement process, Fig.~\ref{fig:rwp change}(a) to (e) show the evolution of $R_{\text{wp}}$ values with iteration number for the five samples, respectively. The agent with four base models demonstrates rapid convergence capability, with efficiency comparable to that of Specialist I and better than that of Specialist II. Both the agent and Specialist I exhibit typical convergence characteristics: a rapid initial decrease followed by stabilization. Overall, both the agent and the two specialists achieve acceptable final $R_{\text{wp}}$ results for all five samples. The agent typically reduces $R_{\text{wp}}$ to a low level within the first 5–10 iterations, among which GPT-5 and MiniMax-m2.7 achieve the fastest convergence and the lowest final values, indicating that the choice of base model significantly influences the refinement decision-making quality of the agent.

It should be noted that both specialists (Specialist I and Specialist II) start refinement from raw data. However, Specialist I, with more experience, implements a more efficient and accurate refinement strategy, reflecting the significant impact of expertise on the refinement process and outcomes. Specialist II, with limited experience, requires many more iterations to reach its own convergence point, highlighting the importance of human expertise in efficiently handling complex parameters. The specialist knowledge base of the Rongzai agent is primarily provided by Specialist I, enabling the agent to learn and inherit more effective refinement strategies. This allows the agent to better assist researchers without prior refinement experience in data analysis.

\subsection{Final refinement quality and time efficiency}

To comprehensively evaluate the refinement performance of the Rongzai agent, the final $R_{\text{wp}}$ values and task times of different methods across five samples are compared. As shown in Table~\ref{tab2}, the Rongzai agent based on MiniMax-m2.7 and GPT-5 achieves the best results on three of the five samples. Specifically, for Zeolite, MiniMax-m2.7 yields the lowest $R_{\text{wp}}$ of 2.88\%, outperforming Specialist I (4.42\%) and Specialist II (4.80\%). For LiNi\(_{0.9}\)Ni\(_{0.05}\)Mn\(_{0.05}\)O\(_{2}\), MiniMax-m2.7 again achieves the best value of 5.06\%, slightly better than Specialist I (5.40\%) and Specialist II (5.70\%). For NaNi\(_{0.33}\)Mn\(_{0.33}\)Fe\(_{0.33}\)O\(_{2}\), GPT-5 obtains the lowest $R_{\text{wp}}$ of 7.60\%, significantly outperforming Specialist I (9.00\%) and Specialist II (9.31\%). On the remaining two samples (Si and LaB\(_6\)), Specialist I maintains the best results, but the agent achieves very close values. These results indicate that the Rongzai agent achieves a performance level comparable to, and in some cases better than, that of an experienced human specialist (Specialist I).

\begin{table}[htbp]
\caption{Comparison of final $R_{\text{wp}}$ (\%) values achieved by the Rongzai agent with different base models and two human specialists across five samples. Bold values indicate the best result for each sample.}\label{tab2}%
\centering
\begin{tabular}{@{}lcccccc@{}}
\toprule
& \multicolumn{4}{@{}c@{}}{Rongzai agent} & \multicolumn{2}{c@{}}{Specialists} \\
\cmidrule(lr){2-5} \cmidrule(lr){6-7}
Samples & MiniMax-m2.7 & GPT-4.1 & GPT-5 & \multicolumn{1}{c}{\shortstack{Scienceone-\\s1-base-ultra}} & $I$ & $II$ \\
\midrule
Si & 4.18 & 18.26 & 3.37 & 3.69 & \textbf{2.94} & 3.53 \\
LaB\textsubscript{6} & 2.65 & 7.15 & 2.86 & 3.71 & \textbf{2.09} & 2.70 \\
Zeolite & \textbf{2.88} & 7.34 & 2.91 & 2.97 & 4.42 & 4.80 \\
LiNi\textsubscript{0.9}Ni\textsubscript{0.05}Mn\textsubscript{0.05}O\textsubscript{2} & \textbf{5.06} & 5.74 & 5.36 & 5.6 & 5.40 & 5.70 \\
NaNi\textsubscript{0.33}Mn\textsubscript{0.33}Fe\textsubscript{0.33}O\textsubscript{2} & 7.85 & 8.64 & \textbf{7.60} & 7.98 & 9.00 & 9.31 \\
\botrule
\end{tabular}
\end{table}

\begin{figure}[htbp]
    \centering
    \includegraphics[width=0.8\textwidth]{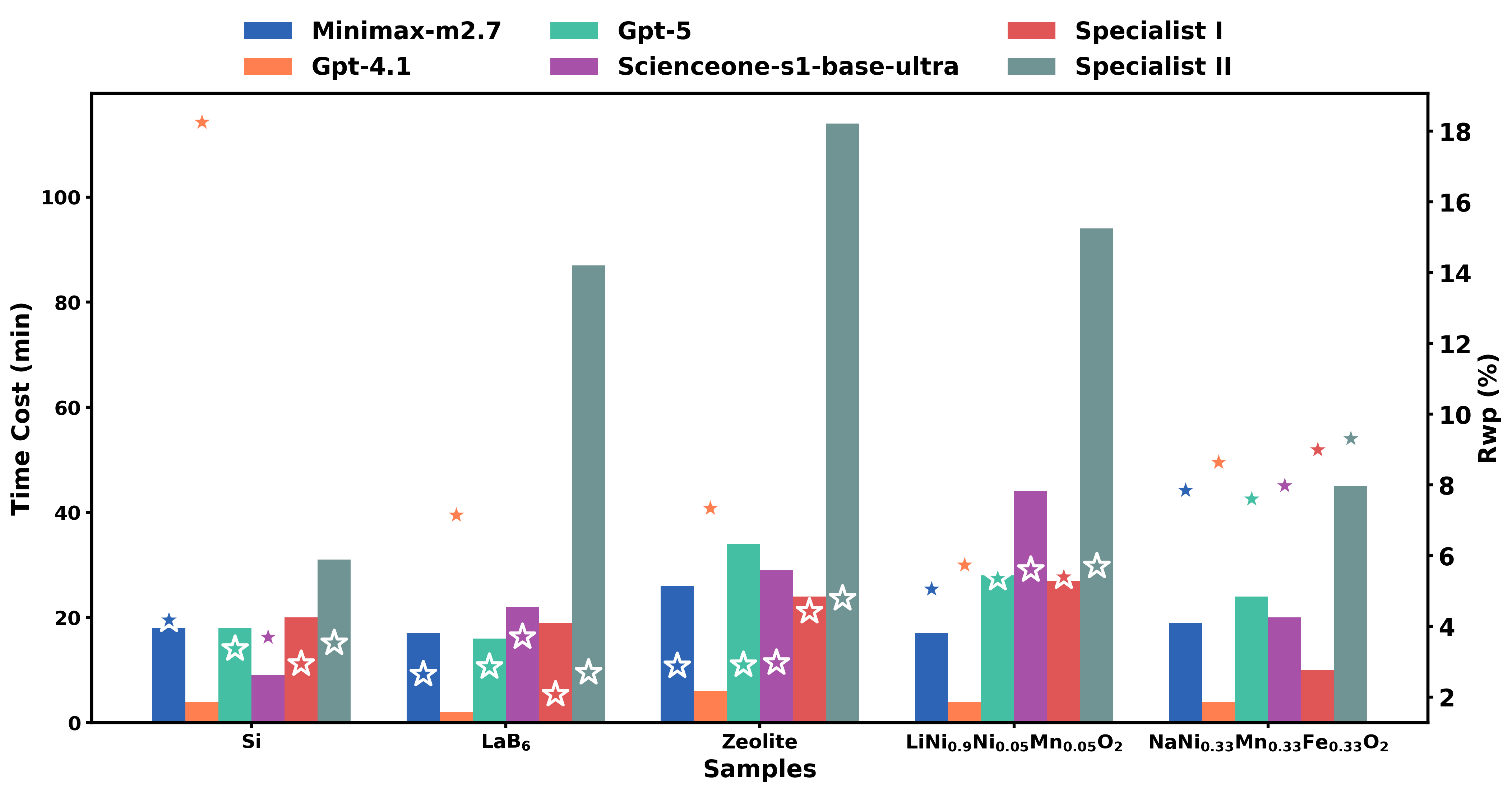}
    \caption{Comparison of task time (bar height, left y-axis) and final $R_{\text{wp}}$ value (star marker, right y-axis) achieved by the Rongzai agent (four base models) and two human specialists (Specialist I, II) across five samples.}
    \label{fig:time_rwp}
\end{figure}

Fig~\ref{fig:time_rwp} illustrates the relationship between task time and final $R_{\text{wp}}$ value, where bar height represents task time and star markers correspond to $R_{\text{wp}}$ values. In terms of time efficiency, the task times of base models (2–44 minutes) are generally significantly lower than those of Specialist II (31–114 minutes). Among them, GPT-5 and MiniMax-m2.7 achieve the best overall performance, maintaining low $R_{\text{wp}}$ values while requiring moderate task times (16–34 minutes). Specialist I requires 10–27 minutes, which falls between the agent and Specialist II, reflecting the trade-off between efficiency and accuracy in manual refinement.

Among the four base models, GPT-5 and MiniMax-m2.7 achieve the best balance between accuracy and efficiency, making them the recommended choices for practical applications. GPT-4.1, while being the most efficient, suffers from insufficient accuracy and is suitable only for time-sensitive scenarios where precision is less critical. Although Scienceone-s1-base-ultra does not achieve the lowest $R_{\text{wp}}$ value for any sample, its refinement results on all five samples are satisfactory, and it even outperforms the human specialists on some samples. Overall, Rongzai agent achieves a refinement quality comparable to that of human specialists, while significantly outperforming Specialist II, who refines from raw data, in terms of efficiency.

\subsection{Visual comparison of fitted patterns}

\begin{figure}[htbp]
    \centering
    \includegraphics[width=0.9\textwidth]{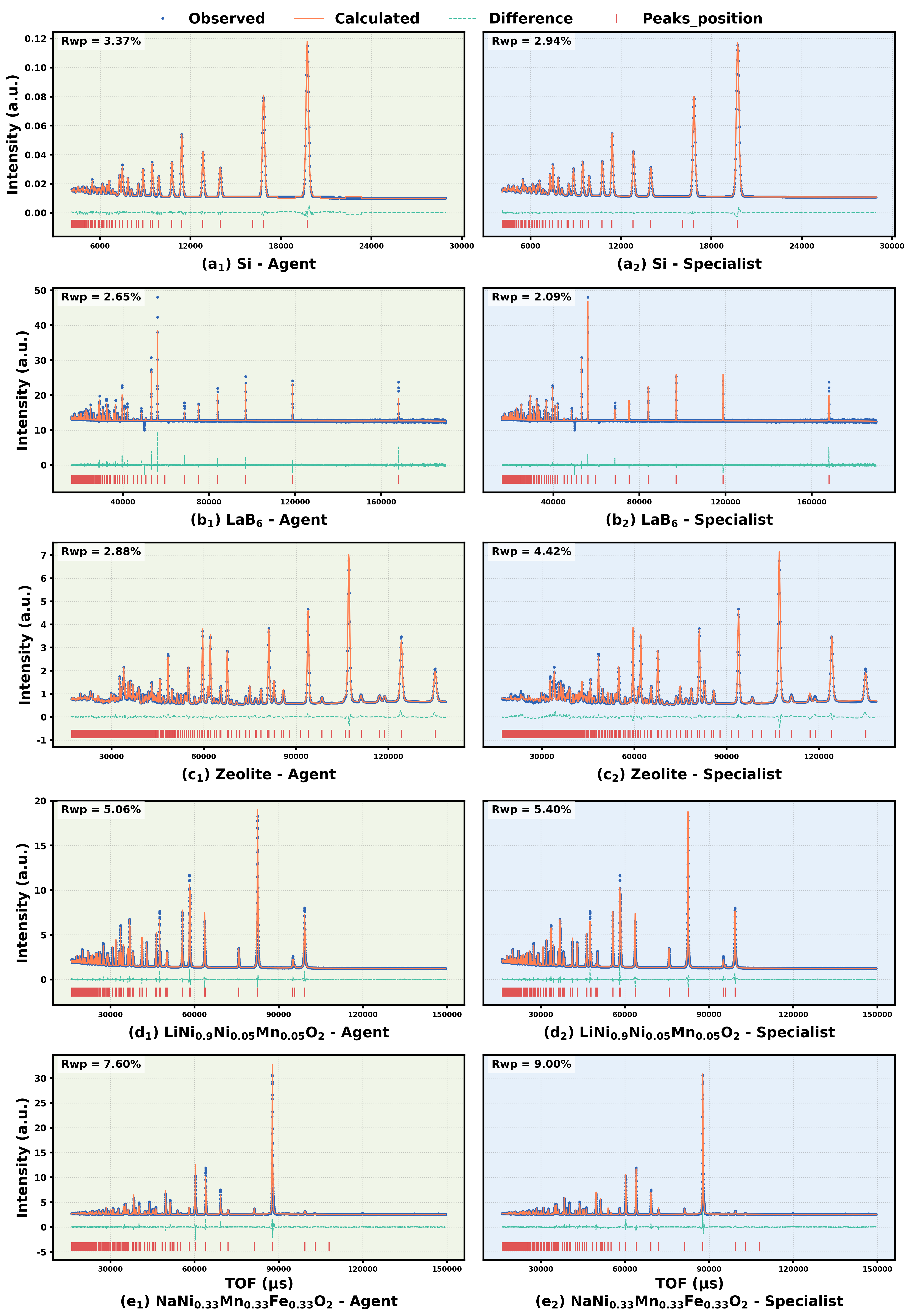}
    \caption{Comparison of fitted neutron powder diffraction patterns for the five samples. Left column: best results achieved by the Rongzai agent (with the corresponding $R_{\text{wp}}$ value). Right column: results obtained by Specialist I (with the corresponding $R_{\text{wp}}$ value). In each panel, observed data (blue circles), calculated profile (orange line), difference curve (green dashed line), and Bragg peak positions (red vertical bars) are shown.}
    \label{fig:refine_plot}
\end{figure}

Given that these parameters cover the majority of neutron powder diffraction data in the absence of magnetic structure, Rongzai agent is well suited to handle most time-of-flight (TOF) ND Rietveld refinement tasks. Although the $R_{\text{wp}}$ is a valid criterion, scientists also assess the acceptability of a refinement by visually inspecting the fit, as shown in Fig.~\ref{fig:refine_plot}. 

Fig.~\ref{fig:refine_plot} compares the fitted patterns of all five samples with the lowest $R_{\text{wp}}$ values achieved by Rongzai agent (using its best-performing base model per sample) and by Specialist I, who possesses extensive refinement experience and provided the expert knowledge base for the agent. For each sample, the left panel shows the agent's best fit, and the right panel shows the specialist's best fit. Overall, the fitted patterns are consistent with the $R_{\text{wp}}$ values reported in Table~\ref{tab2}. For most samples, both Rongzai agent and Specialist I achieve visually satisfactory fits with well-matched peak positions and intensities. For the NaNi\(_{0.33}\)Mn\(_{0.33}\)Fe\(_{0.33}\)O\(_2\) sample (Figs.~\ref{fig:refine_plot}($e_1$) and ($e_2$)), the agent achieves a substantially lower $R_{\text{wp}}$ (7.60\% vs. 9.00\%). However, a closer inspection reveals minor deviations in the agent's fit, primarily in the peak positions within the low-TOF region, where the specialist's fit shows better alignment. Human specialists tend to focus on maintaining as many well-fitted peaks as possible, whereas Rongzai appears to pay less attention to this aspect. Despite these local discrepancies, the agent's overall fit remains acceptable, and the quantitative metric clearly favors the agent. 

The agent's occasional minor discrepancies in local peak positions suggest that further improvement could be achieved by expanding the specialist knowledge base or fine-tuning the LLM with more refined refinement strategies. Nonetheless, the current results demonstrate that Rongzai agent is a reliable and efficient assistant for neutron powder diffraction data analysis, particularly for users without extensive refinement expertise.

In addition, this observation indicates that the Rongzai agent still lacks the specialized knowledge required to handle all aspects of Rietveld refinement reliably. Two possible approaches to address this limitation are: expanding the specialist knowledge base to cover more refinement scenarios, or fine-tuning the LLM by embedding refinement knowledge directly into its model weights. Evidence supporting the second approach is presented in Appendix~\ref{secA1}, which documents a complete refinement process of LaB\(_6\) carried out by Rongzai agent. Because the system prompt is designed to instruct the agent to provide reasoning for each refinement decision, it is possible to distinguish between strategies that are explicitly provided (highlighted in blue) and those that emerge autonomously (highlighted in green). The results show that the agent employs knowledge or strategies that are not provided beforehand. Some of these derive from the LLM itself (e.g., its understanding of the physical meaning and specific effects of instrument parameters, as well as its ability to assess whether parameters are physically reasonable). Others emerge during the refinement process itself, for example, recognizing the need to fix highly correlated parameters to prevent divergence. At the beginning of the refinement, Rongzai agent relies solely on the provided knowledge. However, when it encounters difficulties in reducing $R_{\text{wp}}$, it begins to adopt strategies on its own. Although the foundational knowledge provided is essential for the success of the refinement, this behavior suggests the potential for performing successful refinements without prior domain knowledge, relying entirely on the LLM itself.

\section{Conclusion}\label{sec13}

In this work, we have developed the Rongzai agent, an intelligent autonomous assistant for Rietveld refinement of NPD data, built upon the Open Dr.Sai framework of our team. To the best of our knowledge, this is the first realization of knowledge-driven autonomous refinement using a LLM, where the agent integrates the LLM for task decomposition and decision-making, the GSAS‑II tool as the refinement executor, and a specialist knowledge base to guide refinement strategies. Through a feedback‑driven self‑loop architecture, Rongzai agent achieves fully automated refinement from natural language input to final report generation. Systematic evaluation on five distinct materials yields the following key findings. In terms of refinement quality, Rongzai agent attains the lowest $R_{\text{wp}}$ values on three samples (Zeolite: 2.88\%; LiNi\(_{0.9}\)Ni\(_{0.05}\)Mn\(_{0.05}\)O\(_2\): 5.06\%; NaNi\(_{0.33}\)Mn\(_{0.33}\)Fe\(_{0.33}\)O\(_2\): 7.60\%), substantially outperforming both human specialists. For the remaining two samples, its results are very close to those of the best specialist. Overall, the Rongzai agent represents a significant step toward fully automated and intelligent Rietveld refinement, achieving performance comparable to, and in some cases better than, that of human specialists, while offering superior efficiency. The agent is currently deployed at the CSNS and is open for external use, providing the materials science community with an efficient, user‑friendly tool that tackles a long‑standing challenge in crystallography and exemplifies the potential of AI for Science in accelerating data‑driven discoveries. The successful demonstration of knowledge‑driven decision‑making in an autonomous refinement agent opens new avenues for applying LLMs to complex scientific workflows. Although Rongzai agent performs well quantitatively, it occasionally exhibits local underfitting in specific regions (e.g., low TOF regions), indicating room for improvement in controlling local fitting quality — a limitation that can be addressed in future work by expanding the specialist knowledge base or by fine‑tuning the LLM with domain‑specific refinement knowledge.

\backmatter
\section*{Acknowledgements}
This work was financially supported by the National Key R\&D Program of China (Grant No. 2023YFA1610000); the Strategic Priority Research Program of Chinese Academy of Sciences (Grant No. XDA0480600); and the robotic AI-Scientist platform of Chinese Academy of Sciences.

\section*{Declarations}
\subsection*{Conflicts of interest}
The authors declare that they have no known competing financial interests or personal relationships that could have appeared to influence the work reported in this paper.

\subsection*{Data and code availability}
All data and codes supporting the findings of this study are available from the corresponding authors upon request.

\begin{appendices}
\section[The complete refinement process of Rongzai (GPT-5) for LaB6]{The complete refinement process of Rongzai (GPT-5) for LaB\textsubscript{6}}\label{secA1}
Before the task starting, user attached the files used in refinement: \url{RUN0001239_groupBM.gsa}, \url{groupBM_2025.instprm}, \url{LaB6.cif}.  The the latter reocrd, the blue words means the experiments from the preset specialist knowledge, the green words means the experiments that didn't tell the LLM previously. 

User: Help me to refinement. Use these 3 files.

Rongzai: The user has uploaded .gsa, .instrpm, and .cif files. GSAS-II initialization and \textcolor{blue}{a single empty refinement will be performed directly using these files to generate a .gpx project}. The neutron diffraction data refinement task has been submitted...

\begin{figure}[H]
    \centering
    \includegraphics[width=0.8\textwidth]{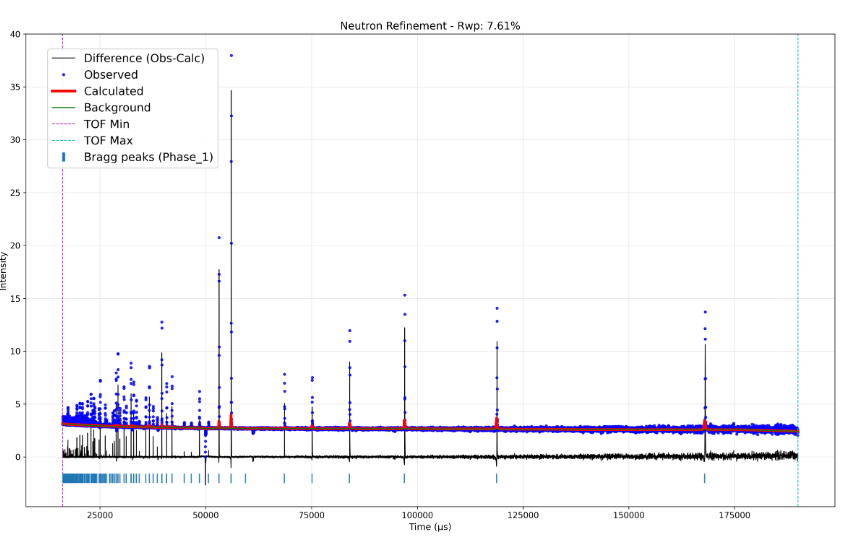}
\end{figure}

Rongzai: The initial refinement only adjusted the scale factor, and the current $R_{\text{wp}}$=7.61\% is still relatively high. \textcolor{blue}{The next step will follow the principle of "only one parameter adjusted at a time": first, only the lattice cell parameters will be opened to correct the peak positions, aiming to improve the overall matching degree; other parameters will remain unchanged}. After the refinement, I will recheck the fitted curves and logs.The neutron diffraction data refinement task has been submitted...

\begin{figure}[H]
    \centering
    \includegraphics[width=0.8\textwidth]{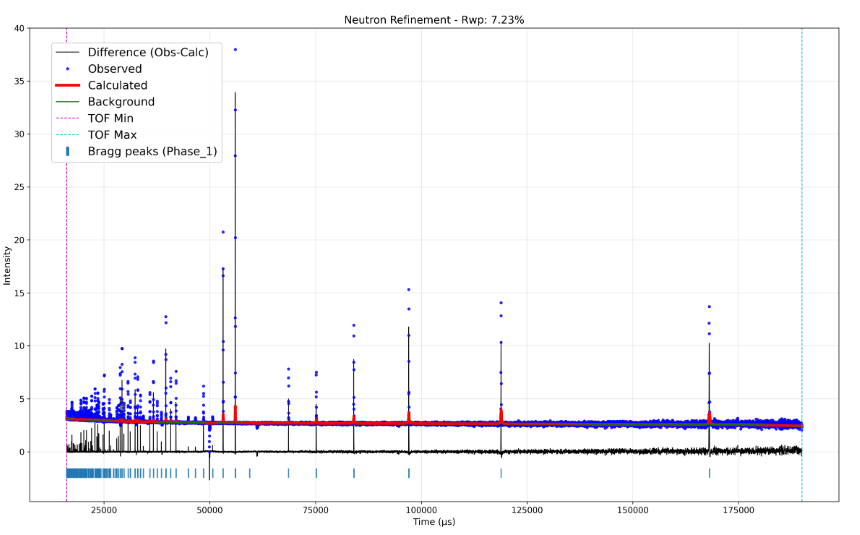}
\end{figure}

Rongzai: The $R_{\text{wp}}$ decreased from 7.61\% to 7.23\%. The peak positions have been improved by refining the cell parameters, but they remain slightly off, and the log still shows a relatively large shift/esd. \textcolor{blue}{Following the strategy of introducing only one new parameter at a time, I kept the scale and cell parameters refined while additionally refining only the instrument alpha to optimize the peak shape/asymmetry, with all other parameters fixed.} After convergence, I will re-examine the fit plot and the log, and refine parameters such as sig and beta sequentially if necessary.

\begin{figure}[H]
    \centering
    \includegraphics[width=0.8\textwidth]{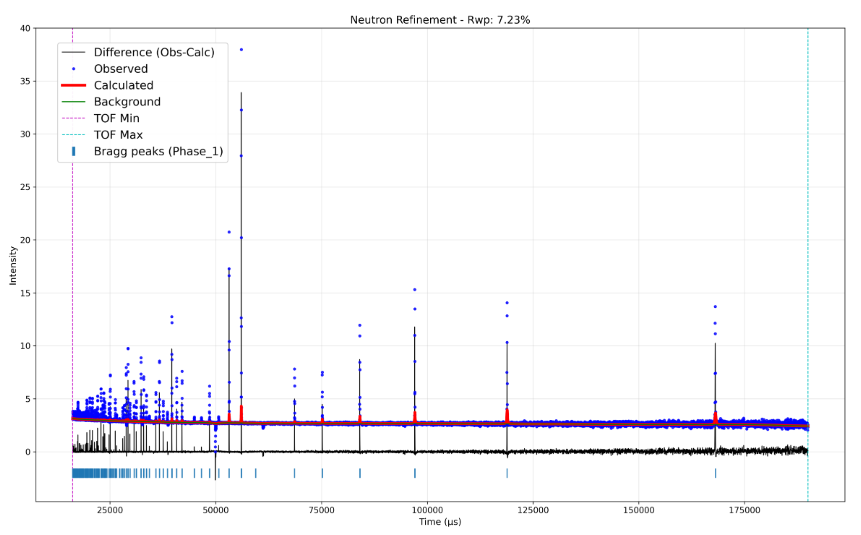}
\end{figure}

Rongzai: \textcolor{blue}{Enabling the instrument alpha in the previous cycle led to SVD failure, so I reverted to the previous version of the GPX file.} The peak positions have already been improved through cell refinement in terms of consistency. \textcolor{blue}{Following the "one new parameter at a time" strategy}, this cycle introduces only the displacement parameter X for B1 (\textcolor{green}{sensitive to intensity}), while \textcolor{blue}{setting the initial microstrain to 0 to eliminate unnecessary peak broadening}. All other parameters remain fixed. After convergence, I will re-examine the fit plot and the log before deciding whether to gradually introduce size or other instrument broadening parameters.

\begin{figure}[H]
    \centering
    \includegraphics[width=0.8\textwidth]{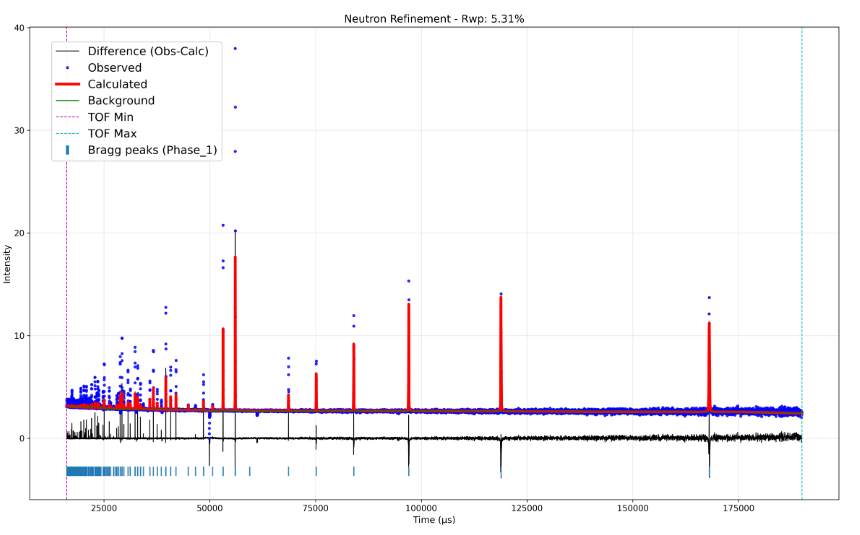}
\end{figure}

Rongzai: The current $R_{\text{wp}}$ is 5.31\%. The refinement of cell parameters and the B1 displacement parameter has significantly improved the peak positions and relative intensities, with mustrain set to 0. However, \textcolor{green}{asymmetry/broadening residuals remain in the peak shapes (see the tail on the low time-of-flight side)}. \textcolor{blue}{Adhering to the strategy of introducing only one new parameter at a time}, while keeping the currently refined parameters (scale, cell, B1:X) open, I will add only the instrument alpha parameter to correct the peak shape asymmetry, with all other parameters fixed. After convergence, if systematic width deviations persist, parameters such as sig or beta will be considered for sequential introduction.

\begin{figure}[H]
    \centering
    \includegraphics[width=0.8\textwidth]{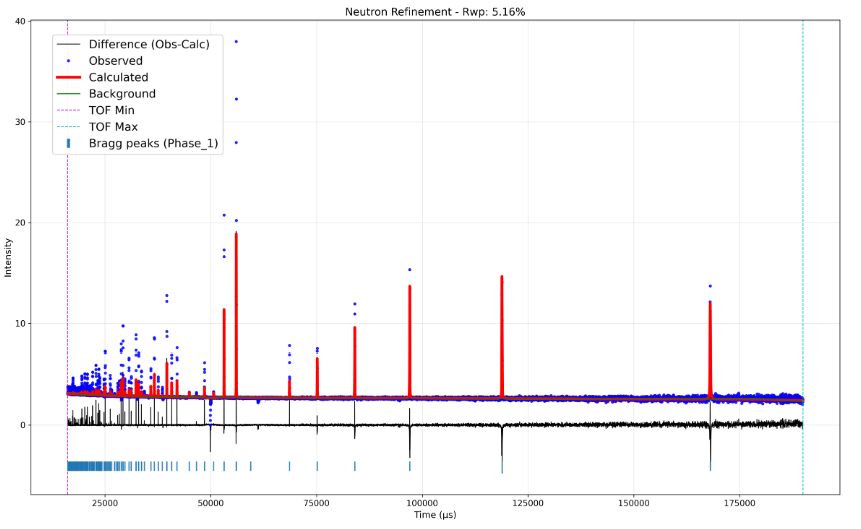}
\end{figure}

Rongzai: The $R_{\text{wp}}$ has decreased to 5.16\%, and the peak positions and relative intensities are generally reasonable. However, systematic deviations remain in the fitted peak widths and tails, and the maximum shift/esd is still relatively large. \textcolor{blue}{Following the strategy of introducing only one new parameter at a time}: while keeping the currently refined parameters (scale, cell, B1 position, and alpha) open, I will add only the instrument sig parameter group (\textcolor{green}{Gaussian broadening}) to better match the peak widths, with all other parameters fixed. After convergence, I will re-examine the fit plot and the log. If Lorentzian residuals persist, the next step will be to introduce the beta group, followed by absorption and background parameters.

\begin{figure}[H]
    \centering
    \includegraphics[width=0.8\textwidth]{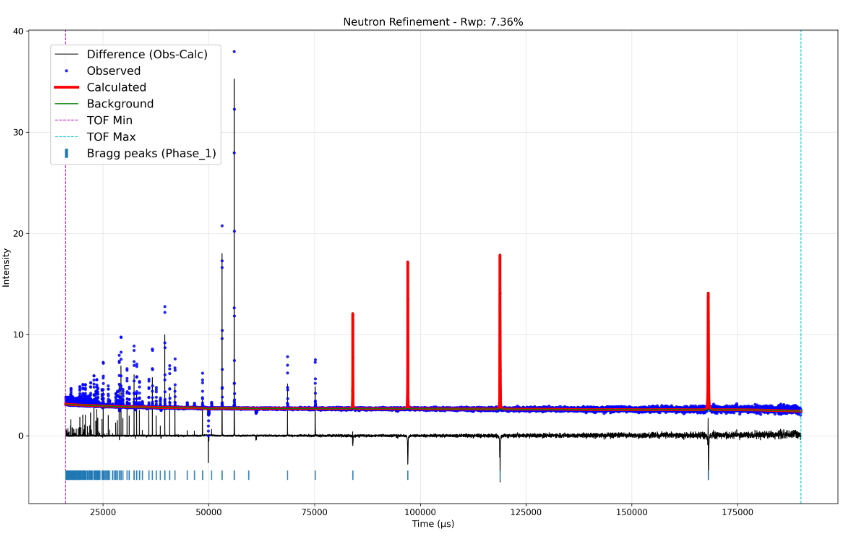}
\end{figure}

Rongzai:"In the previous cycle, enabling the sig group caused the 
$R_{\text{wp}}$ to deteriorate from 5.16\% to 7.36\%, accompanied by high correlations (sig-0/sig-q, sig-1/sig-q) and \textcolor{green}{unphysical parameters (sig-q became negative and unstable)}. Therefore, these changes were rejected, and the refinement was reverted to neutron\_refinement\_4.gpx. To continue optimizing the peak widths and tails while avoiding strong correlations with the sig group, this cycle introduces only \textcolor{green}{the beta group (Lorentzian broadening)}, while keeping the currently refined parameters (scale, cell, B1 position, and alpha) open, aiming to progressively improve the peak shapes. The mustrain parameter remains fixed at 0, and the background is held constant. After convergence, I will examine the fit plot and the log. If the improvement is limited, alternative strategies (such as refining absorption and background) will be considered."

\begin{figure}[H]
    \centering
    \includegraphics[width=0.8\textwidth]{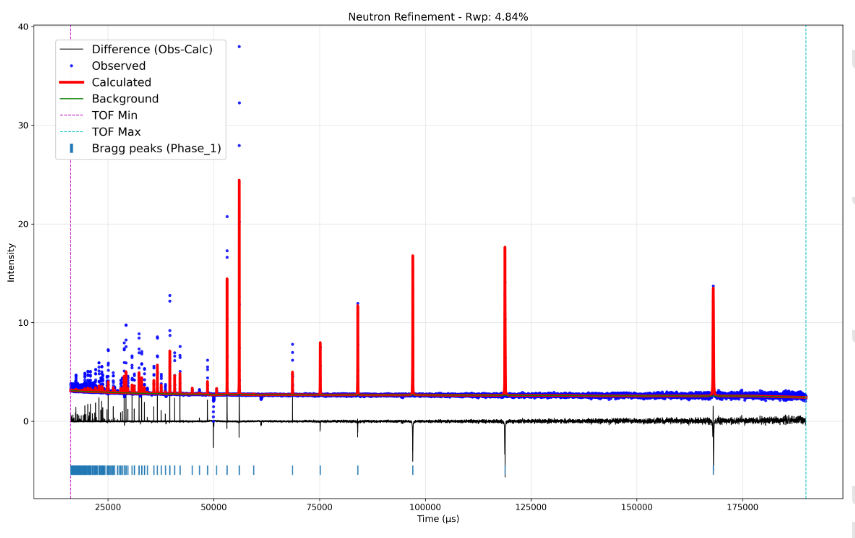}
\end{figure}

Rongzai: "In the previous cycle, after simultaneously refining both alpha and beta, \textcolor{green}{alpha diverged to approximately 1680 and beta-q became negative---physically unreasonable results, despite achieving an $R_{\text{wp}}$ of $4.84\%$. Therefore, these changes were rejected}, and the refinement was reverted to \texttt{neutron\_refinement\_4.gpx}, where alpha remained stable at approximately 0.674. In this cycle, alpha is fixed, and only \textcolor{green}{the beta group (Lorentzian broadening)} is introduced to optimize the peak tails and widths. The cell parameters and B1:X remain refined, while mustrain is fixed at 0 and the background is held constant. After convergence, if the $R_{\text{wp}}$ continues to decrease and the parameters remain physically reasonable, absorption and background refinement will be introduced sequentially. If systematic width residuals persist, the possibility of cautiously refining a minimal number of sig parameters will be reassessed to avoid strong correlations."

\begin{figure}[H]
    \centering
    \includegraphics[width=0.8\textwidth]{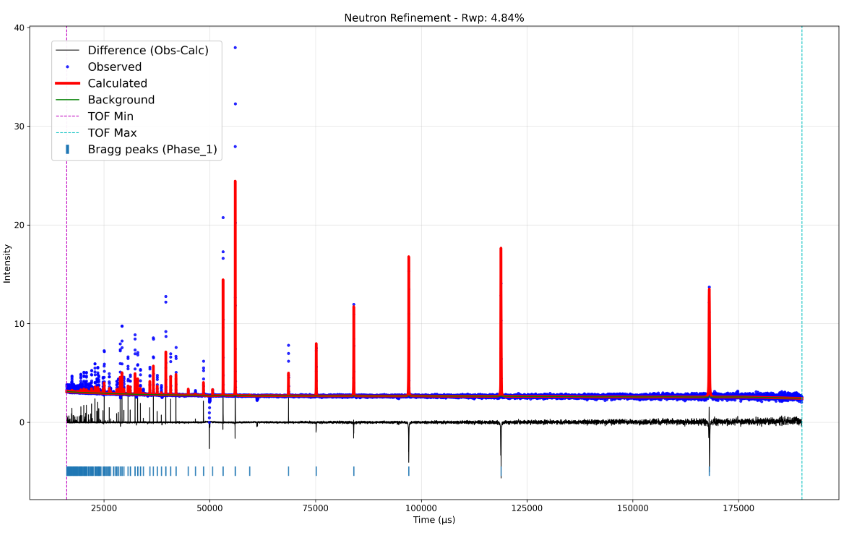}
\end{figure}

Rongzai: "In the previous cycle, alpha was inadvertently kept refined in the current GPX and diverged, resulting in unphysical parameters. Following the established protocol, \textcolor{green}{the refinement was reverted to a stable baseline where alpha was not refined (\texttt{neutron\_refinement\_3.gpx}, $R_{\text{wp}} \approx 5.31\%$}, with cell parameters and B1:X stable and mustrain = 0). In this cycle, adhering to \textcolor{blue}{the one-new-parameter-at-a-time strategy}, only the instrument beta group (\textcolor{green}{Lorentzian broadening}) is introduced to correct the peak tails/widths, while all other parameters remain unchanged (alpha and sig remain off, background and absorption still fixed). After convergence, absorption and background refinement will be introduced sequentially as appropriate."

\begin{figure}[H]
    \centering
    \includegraphics[width=0.8\textwidth]{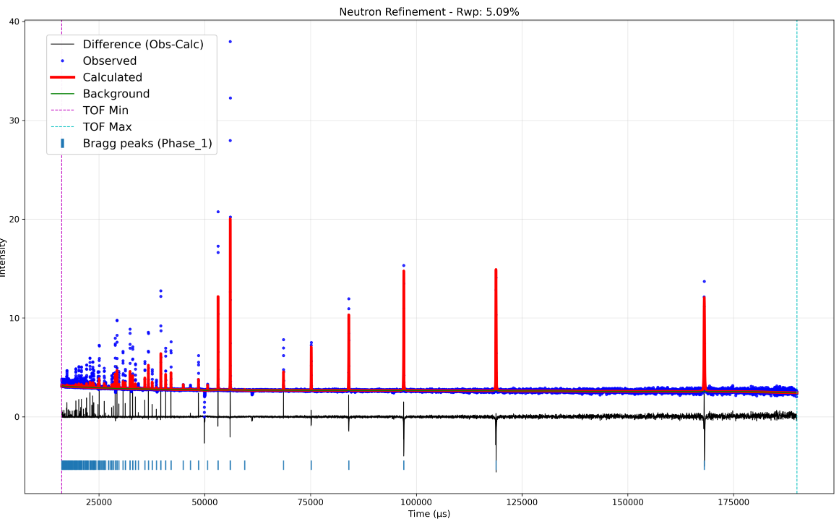}
\end{figure}

Rongzai:"The current baseline is stable: $R_{\text{wp}}$=5.09\%, with reasonable cell parameters and B1:X. The beta group has improved the peak tails, while alpha and sig remain off to avoid divergence or strong correlations. Following the rule of \textcolor{blue}{refining absorption after instrument parameters}, the next step is to introduce only the absorption parameter for refinement, while \textcolor{blue}{temporarily freezing the scale factor to reduce correlation with absorption}. All other refined parameters remain unchanged. After convergence, if the $R_{\text{wp}}$ continues to decrease and the parameters remain physically reasonable, the background coefficients will be refined in the next step for final optimization."

\begin{figure}[H]
    \centering
    \includegraphics[width=0.8\textwidth]{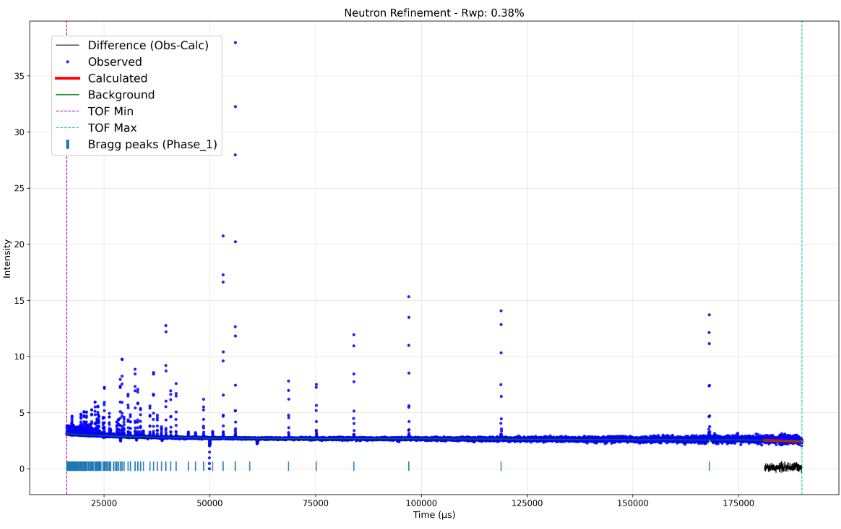}
\end{figure}

Rongzai: ``After absorption was enabled in the previous cycle, obvious unphysical results emerged: \textcolor{green}{the scale factor increased sharply}, the beta parameter became negative and exhibited high correlation with beta-q, and the correlation between scale and absorption exceeded $99\%$. These results were deemed unreliable, and the refinement was reverted accordingly. Following the protocol, the refinement was rolled back to a stable baseline (\texttt{neutron\_refinement\_8.gpx}, $R_{\text{wp}} \approx 5.09\%$, with stable cell parameters, B1:X, and beta; alpha and sig off; mustrain = 0). In this cycle, adhering to\textcolor{blue}{ the one-new-parameter-at-a-time strategy}, only the absorption parameter is refined, with scale explicitly fixed to reduce correlation with absorption. Additionally, \textcolor{blue}{all atomic Uiso values are initialized to 0 to further mitigate the scale---absorption coupling (without introducing new refinable parameters)}. All other parameters remain unchanged. The refinement will be run for two cycles first. After convergence, I will check whether the parameters remain physical and whether the $R_{\text{wp}}$ has decreased.''

\begin{figure}[H]
    \centering
    \includegraphics[width=0.8\textwidth]{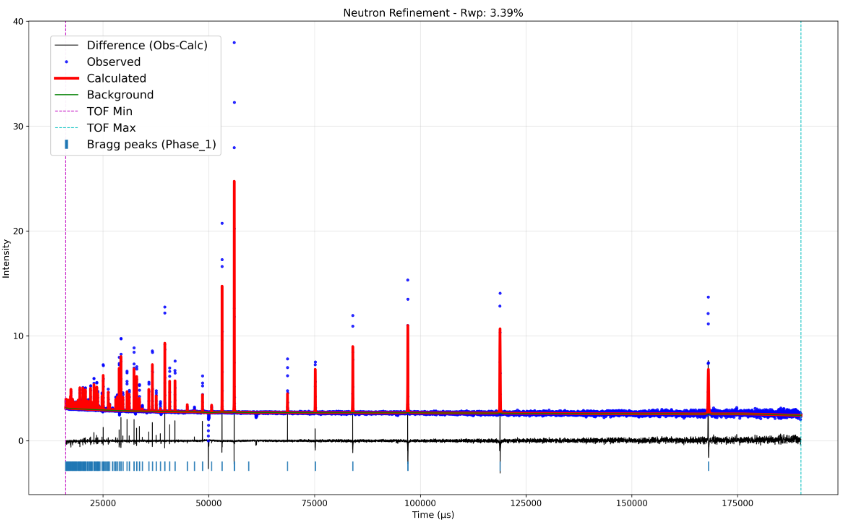}
\end{figure}

Rongzai: ``The current $R_{\text{wp}}$ stands at $3.39\%$, and \textcolor{green}{the parameters are generally physically reasonable (scale $\approx 169$, absorption $\approx 0.155$; the beta group exhibits a mix of positive and negative values but is overall acceptable)}. \textcolor{blue}{Uiso values have been set to zero to reduce correlation with scale and absorption}. The remaining residuals in the fit are primarily slow-varying background features and minor deviations in low-intensity regions. Following the rule that \textcolor{blue}{background should be refined last and must be refined}, this cycle introduces only the background polynomial (Chebyshev type I, with six coefficients) as the final step, while keeping all other refined parameters unchanged (cell parameters, B1 position, beta group, absorption). Scale remains explicitly fixed to avoid coupling with absorption. It is anticipated that this will further reduce the $R_{\text{wp}}$ to below $3\%$. After convergence, I will re-examine the peak positions, peak shapes, and log before deciding whether to fine-tune alpha or restore reasonable Uiso values.''

\begin{figure}[H]
    \centering
    \includegraphics[width=0.8\textwidth]{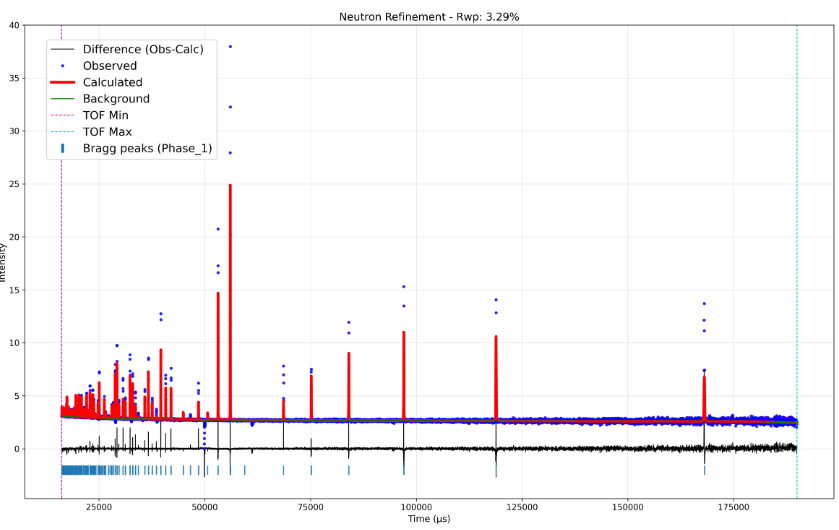}
\end{figure}

Rongzai: ``The current $R_{\text{wp}}$ is $3.29\%$, very close to the target of $3\%$. The peak positions, peak shapes (beta group), and absorption are relatively stable, with the remaining residuals mainly consisting of subtle systematic deviations in intensities. \textcolor{blue}{Following the empirical rule of refining positions before occupancy or thermal parameters for intensity optimization}, the next step introduces only the refinement of atomic Uiso values (opening U for all atoms) to fine-tune the intensity distribution. \textcolor{green}{To avoid strong correlations with absorption and scale, the scale factor is explicitly frozen, while the background, cell parameters, beta group, and absorption remain unchanged}. It is anticipated that this will further reduce the $R_{\text{wp}}$ to below $3\%$. After convergence, I will check whether the Uiso values remain physically reasonable (positive and on the order of $0.005$--$0.05$~\AA$^2$). If anomalies appear, I will revert and consider refining $\alpha$ or slightly increasing the background terms instead.''

\begin{figure}[H]
    \centering
    \includegraphics[width=0.8\textwidth]{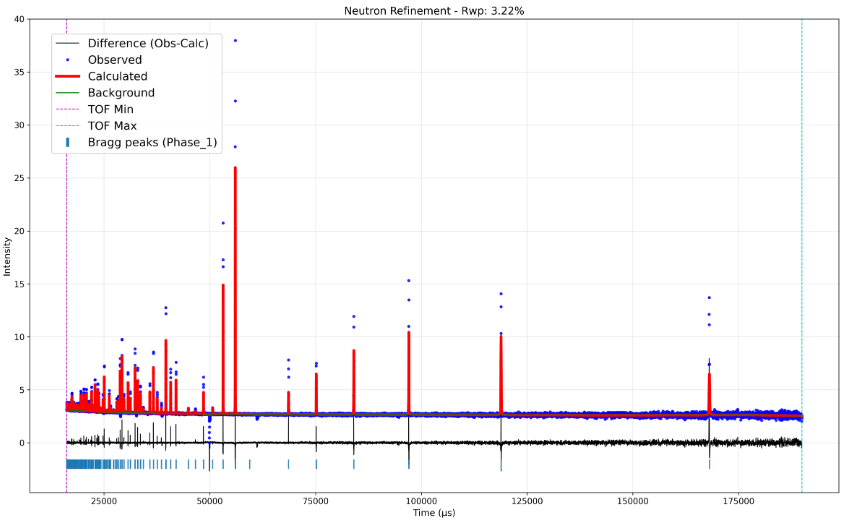}
\end{figure}

Rongzai:"The current $R_{\text{wp}}$ is 3.22\%, very close to the target but still slightly above it. \textcolor{green}{The Uiso values have converged and are physically reasonable}; however, the correlation between scale and absorption exceeds 99\%, necessitating a decoupling strategy. Meanwhile, the fit residuals primarily exhibit slow-varying background features. To address these issues, in this cycle: scale is frozen, absorption remains the only parameter refined among the correlated pair, and the background polynomial order is increased to 8 to better capture the baseline. Cell parameters, the beta group, and all atomic Uiso values remain refined. This is expected to further reduce $R_{\text{wp}}$ to below 3\%. If profile warnings arise or beta parameters become unphysical, I will revert and instead consider fine-tuning the background or refining only beta-0."

\begin{figure}[H]
    \centering
    \includegraphics[width=0.8\textwidth]{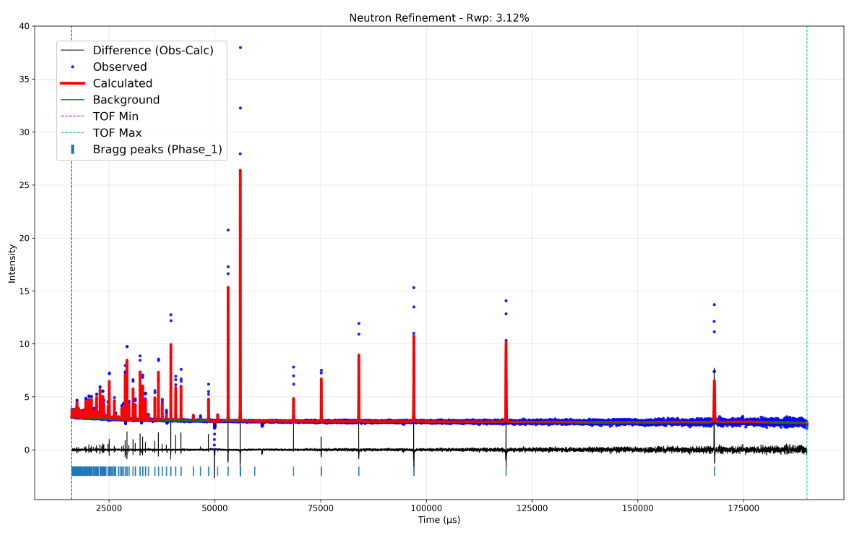}
\end{figure}

Rongzai:"$R_{\text{wp}}$=3.12\%, approaching the target, but \textcolor{green}{slow-varying background residuals persist and intermittent profile warnings (negative widths) have appeared—likely caused by correlations between the beta group and intensity-related parameters}. To steadily reduce $R_{\text{wp}}$ while avoiding unphysical behavior: \textcolor{green}{in this cycle, the beta group, absorption, and scale are frozen (breaking strong correlations)}, retaining only cell parameters, all atomic Uiso values, and the background for refinement. Additionally, the background polynomial order is increased to 10 to better model the slow-varying baseline. A modest improvement is expected, potentially bringing $R_{\text{wp}}$ below 3\%. If the target is still not met, fine-tuning beta-0 or cautiously enabling a minimal number of alpha parameters will be considered next."

\begin{figure}[H]
    \centering
    \includegraphics[width=0.8\textwidth]{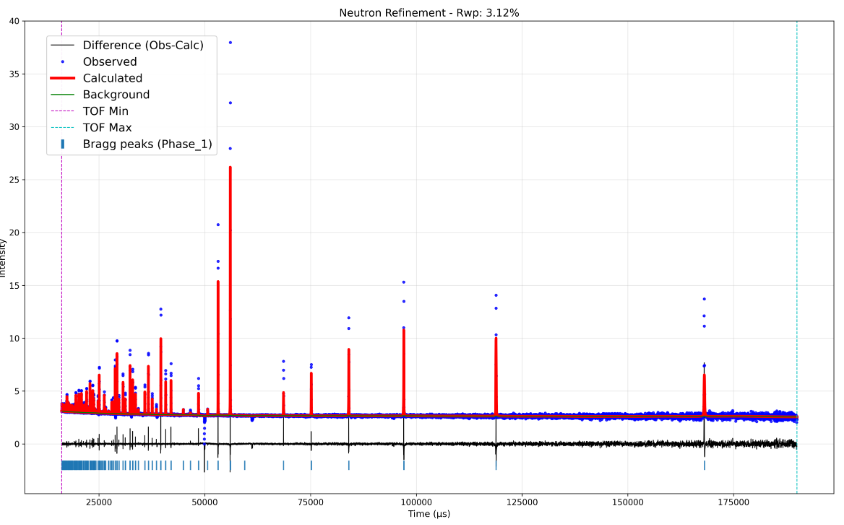}
\end{figure}

Rongzai:"$R_{\text{wp}}$=3.12\%, very close to the target but still slightly above it. To further reduce $R_{\text{wp}}$ without introducing unphysical correlations: in this cycle, only the instrument alpha parameter (\textcolor{green}{describing TOF peak shape asymmetry}) is introduced for refinement, while \textcolor{green}{the beta group, absorption, and scale are explicitly frozen to avoid previously observed high correlations and divergence}. Cell parameters, all atomic Uiso values, and the background (10th order) remain refined, and the number of cycles is moderately increased to 3 to ensure convergence. If alpha becomes unstable or $R_{\text{wp}}$ does not improve, I will revert and instead consider retaining only beta-0 or fine-tuning the background order."

\begin{figure}[H]
    \centering
    \includegraphics[width=0.8\textwidth]{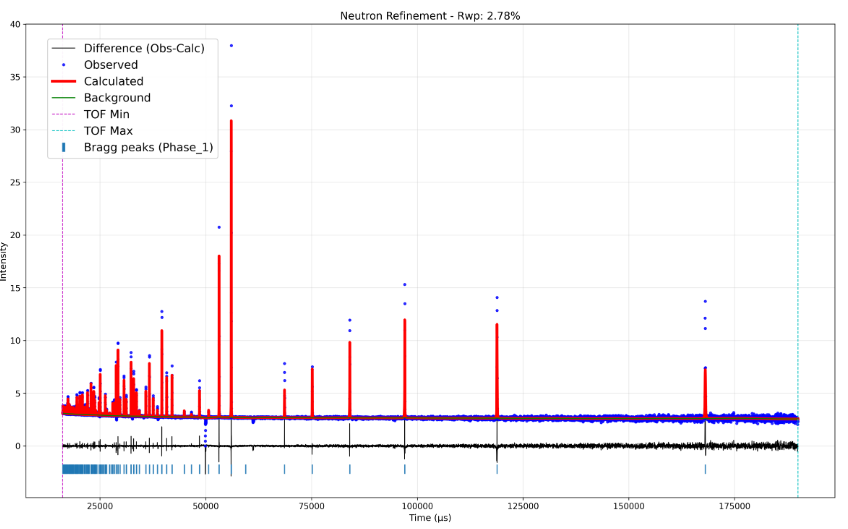}
\end{figure}

Rongzai: ``Refinement completed and report generated. The final results 
show $R_{\text{wp}} = 2.78\%$ and $\chi^2 \approx 2.13 \times 10^4$, with excellent agreement between calculated and experimental patterns in terms of peak positions and profiles, and no SVD warnings or negative peak width errors were encountered. The key refined parameters are as follows: $a = b = c = 4.15663$~\AA, B-site coordinate $x = 0.19966(3)$, $U_{\text{iso}}(\text{La}) = 0.0120(4)$~\AA$^2$, $U_{\text{iso}}(\text{B}) = 0.00345(31)$~\AA$^2$ (both positive and 
physically reasonable), instrument alpha $\approx 9.41$, and the beta 
group ($\beta_0 \approx 0.363$, $\beta_1 \approx 0.763$, $\beta_q \approx -0.568$) remained stable under the current dataset, while mustrain was fixed at 0 and the background using a Chebyshev-I polynomial with 10 coefficients has fully converged. It should be noted 
that scale and absorption remain highly correlated ($>99\%$), but their 
values are stable and do not compromise the current fit quality. The 
final report will include a complete table of refined parameters, the 
evolution of $R_{\text{wp}}$ across refinement cycles, fitted pattern 
comparison plots with residuals, the LST summary and log file excerpts, 
as well as a CIF file and GPX backup (including \texttt{neutron\_refinement\_15.gpx}).''

\end{appendices}

\bibliography{sn-bibliography}


\begin{thebibliography}{19}
\ifx \bisbn   \undefined \def \bisbn  #1{ISBN #1}\fi
\ifx \binits  \undefined \def \binits#1{#1}\fi
\ifx \bauthor  \undefined \def \bauthor#1{#1}\fi
\ifx \batitle  \undefined \def \batitle#1{#1}\fi
\ifx \bjtitle  \undefined \def \bjtitle#1{#1}\fi
\ifx \bvolume  \undefined \def \bvolume#1{\textbf{#1}}\fi
\ifx \byear  \undefined \def \byear#1{#1}\fi
\ifx \bissue  \undefined \def \bissue#1{#1}\fi
\ifx \bfpage  \undefined \def \bfpage#1{#1}\fi
\ifx \blpage  \undefined \def \blpage #1{#1}\fi
\ifx \burl  \undefined \def \burl#1{\textsf{#1}}\fi
\ifx \doiurl  \undefined \def \doiurl#1{\url{https://doi.org/#1}}\fi
\ifx \betal  \undefined \def \betal{\textit{et al.}}\fi
\ifx \binstitute  \undefined \def \binstitute#1{#1}\fi
\ifx \binstitutionaled  \undefined \def \binstitutionaled#1{#1}\fi
\ifx \bctitle  \undefined \def \bctitle#1{#1}\fi
\ifx \beditor  \undefined \def \beditor#1{#1}\fi
\ifx \bpublisher  \undefined \def \bpublisher#1{#1}\fi
\ifx \bbtitle  \undefined \def \bbtitle#1{#1}\fi
\ifx \bedition  \undefined \def \bedition#1{#1}\fi
\ifx \bseriesno  \undefined \def \bseriesno#1{#1}\fi
\ifx \blocation  \undefined \def \blocation#1{#1}\fi
\ifx \bsertitle  \undefined \def \bsertitle#1{#1}\fi
\ifx \bsnm \undefined \def \bsnm#1{#1}\fi
\ifx \bsuffix \undefined \def \bsuffix#1{#1}\fi
\ifx \bparticle \undefined \def \bparticle#1{#1}\fi
\ifx \barticle \undefined \def \barticle#1{#1}\fi
\bibcommenthead
\ifx \bconfdate \undefined \def \bconfdate #1{#1}\fi
\ifx \botherref \undefined \def \botherref #1{#1}\fi
\ifx \url \undefined \def \url#1{\textsf{#1}}\fi
\ifx \bchapter \undefined \def \bchapter#1{#1}\fi
\ifx \bbook \undefined \def \bbook#1{#1}\fi
\ifx \bcomment \undefined \def \bcomment#1{#1}\fi
\ifx \oauthor \undefined \def \oauthor#1{#1}\fi
\ifx \citeauthoryear \undefined \def \citeauthoryear#1{#1}\fi
\ifx \endbibitem  \undefined \def \endbibitem {}\fi
\ifx \bconflocation  \undefined \def \bconflocation#1{#1}\fi
\ifx \arxivurl  \undefined \def \arxivurl#1{\textsf{#1}}\fi
\csname PreBibitemsHook\endcsname

\bibitem[\protect\citeauthoryear{Haberl et~al.}{}]{haberl_advancing_2023}
\begin{botherref}
\oauthor{\bsnm{Haberl}, \binits{B.}},
\oauthor{\bsnm{Guthrie}, \binits{M.}},
\oauthor{\bsnm{Boehler}, \binits{R.}}:
Advancing neutron diffraction for accurate structural measurement of light
  elements at megabar pressures
\textbf{13}(1),
4741
\end{botherref}
\endbibitem

\bibitem[\protect\citeauthoryear{Pomjakushin
  et~al.}{}]{pomjakushin_topological_2023}
\begin{botherref}
\oauthor{\bsnm{Pomjakushin}, \binits{V.}},
\oauthor{\bsnm{Plokhikh}, \binits{I.}},
\oauthor{\bsnm{White}, \binits{J.}},
\oauthor{\bsnm{Fujishiro}, \binits{Y.}},
\oauthor{\bsnm{Kanazawa}, \binits{N.}},
\oauthor{\bsnm{Tokura}, \binits{Y.}},
\oauthor{\bsnm{Pomjakushina}, \binits{E.}}:
Topological magnetic structures in {MnGe}: Neutron diffraction and symmetry
  analysis
\textbf{107}(2),
024410
\end{botherref}
\endbibitem

\bibitem[\protect\citeauthoryear{Vershinina
  et~al.}{}]{vershinina_comparative_2023}
\begin{botherref}
\oauthor{\bsnm{Vershinina}, \binits{T.}},
\oauthor{\bsnm{Samoylova}, \binits{N.Y.}},
\oauthor{\bsnm{Sumnikov}, \binits{S.}},
\oauthor{\bsnm{Balagurov}, \binits{A.}},
\oauthor{\bsnm{Palacheva}, \binits{V.}},
\oauthor{\bsnm{Golovin}, \binits{I.}}:
Comparative study of structures and phase transitions in fe-(31-35) at\% ga
  alloys by in situ neutron diffraction
\textbf{934},
167967
\end{botherref}
\endbibitem

\bibitem[\protect\citeauthoryear{De~Wolff}{}]{de_wolff_determination_1957}
\begin{botherref}
\oauthor{\bsnm{De~Wolff}, \binits{P.}}:
On the determination of unit-cell dimensions from powder diffraction patterns
\textbf{10}(9),
590--595
\end{botherref}
\endbibitem

\bibitem[\protect\citeauthoryear{Mighell and
  Santoro}{}]{mighell_geometrical_1975}
\begin{botherref}
\oauthor{\bsnm{Mighell}, \binits{A.t.}},
\oauthor{\bsnm{Santoro}, \binits{A.}}:
Geometrical ambiguities in the indexing of powder patterns
\textbf{8}(3),
372--374
\end{botherref}
\endbibitem

\bibitem[\protect\citeauthoryear{Rodriguez-Carvajal}{}]{rodriguez-carvajal_recent_2001}
\begin{botherref}
\oauthor{\bsnm{Rodriguez-Carvajal}, \binits{J.}}:
Recent developments of the program {FULLPROF}, commission on powder diffraction
\textbf{26}
\end{botherref}
\endbibitem

\bibitem[\protect\citeauthoryear{Toby and Von~Dreele}{}]{toby_gsas-ii_2013}
\begin{botherref}
\oauthor{\bsnm{Toby}, \binits{B.H.}},
\oauthor{\bsnm{Von~Dreele}, \binits{R.B.}}:
{GSAS}-{II}: the genesis of a modern open-source all purpose crystallography
  software package
\textbf{46}(2),
544--549
\end{botherref}
\endbibitem

\bibitem[\protect\citeauthoryear{Toby}{}]{toby_simple_2024}
\begin{botherref}
\oauthor{\bsnm{Toby}, \binits{B.}}:
A simple solution to the rietveld refinement recipe problem
\textbf{57}(1),
175--180
\end{botherref}
\endbibitem

\bibitem[\protect\citeauthoryear{Young}{}]{young_rietveld_1993}
\begin{botherref}
\oauthor{\bsnm{Young}, \binits{R.A.}}:
The Rietveld Method
vol. 5.
International union of crystallography
\end{botherref}
\endbibitem

\bibitem[\protect\citeauthoryear{Biwer et~al.}{}]{biwer_spotlight_2025}
\begin{botherref}
\oauthor{\bsnm{Biwer}, \binits{C.M.}},
\oauthor{\bsnm{Feng}, \binits{Z.}},
\oauthor{\bsnm{Finstad}, \binits{D.}},
\oauthor{\bsnm{{McDonnell}}, \binits{M.}},
\oauthor{\bsnm{Knezevic}, \binits{M.}},
\oauthor{\bsnm{{McKerns}}, \binits{M.}},
\oauthor{\bsnm{Savage}, \binits{D.J.}},
\oauthor{\bsnm{Vogel}, \binits{S.C.}}:
Spotlight: efficient automated global optimization in rietveld analysis of
  diffraction data
\textbf{15}(1),
8358
\end{botherref}
\endbibitem

\bibitem[\protect\citeauthoryear{Tian et~al.}{}]{tian_srrietveld_2013}
\begin{botherref}
\oauthor{\bsnm{Tian}, \binits{P.}},
\oauthor{\bsnm{Zhou}, \binits{W.}},
\oauthor{\bsnm{Liu}, \binits{J.}},
\oauthor{\bsnm{Shang}, \binits{Y.}},
\oauthor{\bsnm{Farrow}, \binits{C.}},
\oauthor{\bsnm{Juhás}, \binits{P.}},
\oauthor{\bsnm{Billinge}, \binits{S.}}:
{SrRietveld}: a program for automating rietveld refinements for high-throughput
  powder diffraction studies
\textbf{46}(1),
255--258
\end{botherref}
\endbibitem

\bibitem[\protect\citeauthoryear{Cui et~al.}{}]{cui_autofp_2015}
\begin{botherref}
\oauthor{\bsnm{Cui}, \binits{X.}},
\oauthor{\bsnm{Feng}, \binits{Z.}},
\oauthor{\bsnm{Jin}, \binits{Y.}},
\oauthor{\bsnm{Cao}, \binits{Y.}},
\oauthor{\bsnm{Deng}, \binits{D.}},
\oauthor{\bsnm{Chu}, \binits{H.}},
\oauthor{\bsnm{Cao}, \binits{S.}},
\oauthor{\bsnm{Dong}, \binits{C.}},
\oauthor{\bsnm{Zhang}, \binits{J.}}:
{AutoFP}: a {GUI} for highly automated rietveld refinement using an expert
  system algorithm based on {FullProf}
\textbf{48}(5),
1581--1586
\end{botherref}
\endbibitem

\bibitem[\protect\citeauthoryear{Feng et~al.}{}]{feng_method_2019}
\begin{botherref}
\oauthor{\bsnm{Feng}, \binits{Z.}},
\oauthor{\bsnm{Hou}, \binits{Q.}},
\oauthor{\bsnm{Zheng}, \binits{Y.}},
\oauthor{\bsnm{Ren}, \binits{W.}},
\oauthor{\bsnm{Ge}, \binits{J.-Y.}},
\oauthor{\bsnm{Li}, \binits{T.}},
\oauthor{\bsnm{Cheng}, \binits{C.}},
\oauthor{\bsnm{Lu}, \binits{W.}},
\oauthor{\bsnm{Cao}, \binits{S.}},
\oauthor{\bsnm{Zhang}, \binits{J.}}:
Method of artificial intelligence algorithm to improve the automation level of
  rietveld refinement
\textbf{156},
310--314
\end{botherref}
\endbibitem

\bibitem[\protect\citeauthoryear{Aimi and {Kenjiro
  Fujimoto}}{}]{aimi_development_2019}
\begin{botherref}
\oauthor{\bsnm{Aimi}, \binits{A.}},
\oauthor{\bsnm{{Kenjiro Fujimoto}}}:
Development of an automatic, high-throughput structural refinement method using
  rietveld analysis
\textbf{22}(1),
35--41
\end{botherref}
\endbibitem

\bibitem[\protect\citeauthoryear{Ozaki et~al.}{}]{ozaki_automated_2020}
\begin{botherref}
\oauthor{\bsnm{Ozaki}, \binits{Y.}},
\oauthor{\bsnm{Suzuki}, \binits{Y.}},
\oauthor{\bsnm{Hawai}, \binits{T.}},
\oauthor{\bsnm{Saito}, \binits{K.}},
\oauthor{\bsnm{Onishi}, \binits{M.}},
\oauthor{\bsnm{Ono}, \binits{K.}}:
Automated crystal structure analysis based on blackbox optimisation
\textbf{6}(1),
75
\end{botherref}
\endbibitem

\bibitem[\protect\citeauthoryear{Ai et~al.}{2026}]{ai_scivisagentbench_2026}
\begin{botherref}
\oauthor{\bsnm{Ai}, \binits{K.}},
\oauthor{\bsnm{Miao}, \binits{H.}},
\oauthor{\bsnm{Tang}, \binits{K.}}, et al.:
{SciVisAgentBench}: A benchmark for evaluating scientific data analysis and
  visualization agents.
arXiv preprint arXiv:2603.29139
(2026)
\end{botherref}
\endbibitem

\bibitem[\protect\citeauthoryear{Wu et~al.}{2026}]{wu_omniflow_2026}
\begin{botherref}
\oauthor{\bsnm{Wu}, \binits{H.}},
\oauthor{\bsnm{Zhang}, \binits{Y.}},
\oauthor{\bsnm{Gao}, \binits{Y.}}, et al.:
{OMNIFLOW}: A physics-grounded multimodal agent for generalized scientific
  reasoning.
arXiv preprint arXiv:2603.15797
(2026)
\end{botherref}
\endbibitem

\bibitem[\protect\citeauthoryear{Li et~al.}{}]{li_constructing_2026}
\begin{botherref}
\oauthor{\bsnm{Li}, \binits{C.}},
\oauthor{\bsnm{Ji}, \binits{W.}},
\oauthor{\bsnm{Xu}, \binits{D.}},
\oauthor{\bsnm{Xu}, \binits{W.}},
\oauthor{\bsnm{Qiu}, \binits{T.}},
\oauthor{\bsnm{Qiao}, \binits{D.}},
\oauthor{\bsnm{Zhan}, \binits{X.}},
\oauthor{\bsnm{Miao}, \binits{P.}},
\oauthor{\bsnm{Sun}, \binits{Y.}},
\oauthor{\bsnm{Chen}, \binits{Z.}},
\oauthor{\bsnm{Zhang}, \binits{Q.}}:
Constructing a stabilized interface in ultra-high nickel single-crystal
  {LiNi}0.90co0.05mn0.05o2 by a long-time molten-salt route
\textbf{10}(4),
02319
\end{botherref}
\endbibitem

\bibitem[\protect\citeauthoryear{Xie et~al.}{}]{xie_probing_2018}
\begin{botherref}
\oauthor{\bsnm{Xie}, \binits{Y.}},
\oauthor{\bsnm{Xu}, \binits{G.-L.}},
\oauthor{\bsnm{Che}, \binits{H.}},
\oauthor{\bsnm{Wang}, \binits{H.}},
\oauthor{\bsnm{Yang}, \binits{K.}},
\oauthor{\bsnm{Yang}, \binits{X.}},
\oauthor{\bsnm{Guo}, \binits{F.}},
\oauthor{\bsnm{Ren}, \binits{Y.}},
\oauthor{\bsnm{Chen}, \binits{Z.}},
\oauthor{\bsnm{Amine}, \binits{K.}},
\oauthor{\bsnm{Ma}, \binits{Z.-F.}}:
Probing thermal and chemical stability of {NaxNi}1/3fe1/3mn1/3o2 cathode
  material toward safe sodium-ion batteries
\textbf{30}(15),
4909--4918
\doiurl{10.1021/acs.chemmater.8b00047}
\end{botherref}
\endbibitem

\end{thebibliography}

\end{document}